\documentclass[prl,aps,twocolumn]{revtex4}

\usepackage{revsymb}
\usepackage{graphicx}
\usepackage{dcolumn}
\usepackage{bm}

\usepackage{epstopdf}

\begin{document}

\title{Charge control in InP/GaInP single quantum dots embedded in Schottky diodes}

\author{O.~D.~D.~Couto~Jr.}
\email{o.couto@sheffield.ac.uk}
\author{J.~Puebla}
\author{E.~A.~Chekhovich,~I.~J.~Luxmoore,~C.~J.~Elliott}
\author{N.~Babazadeh}
\author{M.~S.~Skolnick}
\author{A.~I.~Tartakovskii}
\email{a.tartakovskii@sheffield.ac.uk}
\affiliation{Department of Physics and Astronomy, University of Sheffield, Sheffield S3 7RH, UK}
\author{A.~B.~Krysa}
\affiliation{Department of Electronic and Electrical Engineering, University of Sheffield, Sheffield S1 3JD, UK}

\date{\today}

\begin{abstract}
We demonstrate control by applied electric field of the charge states in single self-assembled InP quantum dots placed in GaInP Schottky structures grown by metalorganic vapor phase epitaxy. This has been enabled by growth optimization leading to suppression of formation of large dots uncontrollably accumulating charge.  Using bias- and polarization-dependent micro-photoluminescence, we identify the exciton multi-particle states and carry out a systematic study of the neutral exciton state dipole moment and polarizability. This analysis allows for the characterization of the exciton wavefunction properties at the single dot level for this type of quantum dots. Photocurrent measurements allow further characterization of exciton properties by electrical means, opening new possibilities for resonant excitation studies for such system.
\end{abstract}

\maketitle

\section{I. Introduction}

Experiments on individual quantum dots (QDs) have revealed a wealth of effects due to strong confinement and the resulting isolation of charge-carriers in these nano-structures from the surrounding bulk material~\cite{qbit_book}.  Self-assembled QDs have been widely researched for applications ranging from single-photon emitters~\cite{qbit_book,Michler} to spin qubits for quantum information processing~\cite{qbit_book,Atature,Ramsay}. In particular, InP/GaInP QDs provide stable single-photon sources in the red spectral range~\cite{Zwiller, Richter}, where current silicon-based single-photon detectors have their highest detection efficiency. Recent studies have also shown intriguing nuclear spin phenomena in InP/GaInP QDs grown by MOVPE. In optically pumped QDs, record high degrees of nuclear spin polarization $\approx 65\%$~\cite{Skiba,Chekhovich1} and ultra-long nuclear depolarization times up to $5000\mathrm{s}$ have been observed~\cite{Chekhovich3}. A direct measurement of the hole hyperfine interaction in semiconductors has also been demonstrated~\cite{Chekhovich2}, placing these dots in the context of the intensively pursued research into QD-based spin qubits~\cite{qbit_book}.

Nevertheless, the use of InP dots for spin studies encounters the following major challenge: InP/GaInP samples  commonly contain multi-modal distributions of QD sizes, consisting of lower energy ($1.6-1.7~\mathrm{eV}$ at 10~K) large QDs and higher energy ($1.7-1.9~\mathrm{eV}$ at 10~K) small QDs \cite{Schulz,Persson}. Although these reproducibly grown samples allow access to individual small QDs in the high energy range, their properties are uncontrollably influenced by interactions with high density large QDs.  These large QDs have been shown to accumulate high numbers of charges at low temperatures~\cite{Hessman}, leading to charge instability and additional spin relaxation pathways in the neighboring small QDs. The presence of large dots is the most likely reason for a certain time delay, in comparison to InGaAs/GaAs structures, for the achievement of effective charge control in InP single QDs placed in Schottky diodes. This is now realized in our work, after the growth of QDs with a single-mode size distribution has been achieved. This essential step enabled realization of charge-tunable InP QDs for future studies of charge-controlled few spin nano-systems.

In this paper, we report on control by electric field of exciton charge states in individual InP dots by placing them in the intrinsic region of \emph{n-i}-Schottky diode structures. The optimized growth using low-pressure metalorganic vapor phase epitaxy (MOVPE) enabled us to avoid formation of high densities of large highly charged QDs, leading to optimized samples containing only small QDs with densities below $10^9\mathrm{cm}^{-2}$. From bias and polarization-dependent analysis of the photoluminescence (PL), multi-particle excitonic complexes could be observed and identified  as the neutral ($X_0$), singly ($X^{-1}$), doubly ($X^{-2}$) and even triply ($X^{-3}$) negatively charged excitons. Binding energies for the $X^{-1}$ are demonstrated to range from 4 to 7~meV. We probe the PL bias dependence of a relatively high number of individual dots, which allows for a general characterization of the electron-hole permanent dipole moment and polarizability of this system. From the dipole moment analysis, we demonstrate that for InP/GaInP QDs the electron-hole alignment along the growth direction is generally opposite to what is usually observed for InGaAs/GaAs QDs. From the polarizability study, we characterize the lateral extent of the exciton wavefunction in the QD plane and the hole wavefunction extension along the growth direction. Complementary to PL measurements, we carry out resonant excitation experiments, where photon absorption by the dot is measured using photocurrent (PC) technique, opening perspectives to manipulate the electron and hole lifetimes for application in resonant coherent spin control measurements \cite{Ramsay,Ramsay2}.

The manuscript is organized as follows. We start in Sec. II with a description of the samples structure and the experiments.
The experimental results are presented in Sec. III, where in subsection A we discuss the QD growth optimization procedure necessary to obtain a uniform distribution of QD sizes. Subsections C and D are devoted to single QD characterization and statistical analysis of QD properties performed in a large ensemble of QDs, respectively. In subsection D we present single QD characterization by resonant photocurrent spectroscopy. Section IV summarizes the main conclusions of this work.

\section{II. Experimental}

The sample growth was performed in a horizontal flow quartz reactor using low-pressure MOVPE on (100) \textit{n}-type GaAs substrates misorientated  by $3^{\circ}$ towards $\langle111\rangle$. The growth temperature of the $\mathrm{GaAs}$ buffer and bottom $\mathrm{Ga_{0.5}In_{0.5}P}$ layer was $700\mathrm{^{\circ}}$C. Before proceeding to the deposition of $\mathrm{InP}$ and the $\mathrm{Ga_{0.5}In_{0.5}P}$ capping layer, the wafer was cooled to $650\mathrm{^{\circ}}$C. The grown $\mathrm{GaInP}$ layers were nominally lattice matched to $\mathrm{GaAs}$. A low $\mathrm{InP}$ growth rate of $1.1$\AA$\mathrm{/s}$ and deposition time of 3 seconds was chosen for the optimized samples. Two different Schottky diode structures were analyzed (see inset on Fig.~\ref{Fig2} for a detailed description). Sample A consisted of a QD layer grown on top of a 40 nm thick \textit{i}-GaInP layer above the \textit{n}-doped GaInP region. Capping was performed with a 160~nm-thickness \textit{i}-GaInP layer only. Sample B consisted of a QD layer also grown 40 nm above the \textit{n}-doped GaInP region, but capped by a sequence of undoped GaInP/AlGaInP/GaInP layers with thicknesses of 85, 25, and 50~nm, respectively, in order to create a blocking barrier for holes.

\begin{figure}
\includegraphics[width=7.5cm,clip]{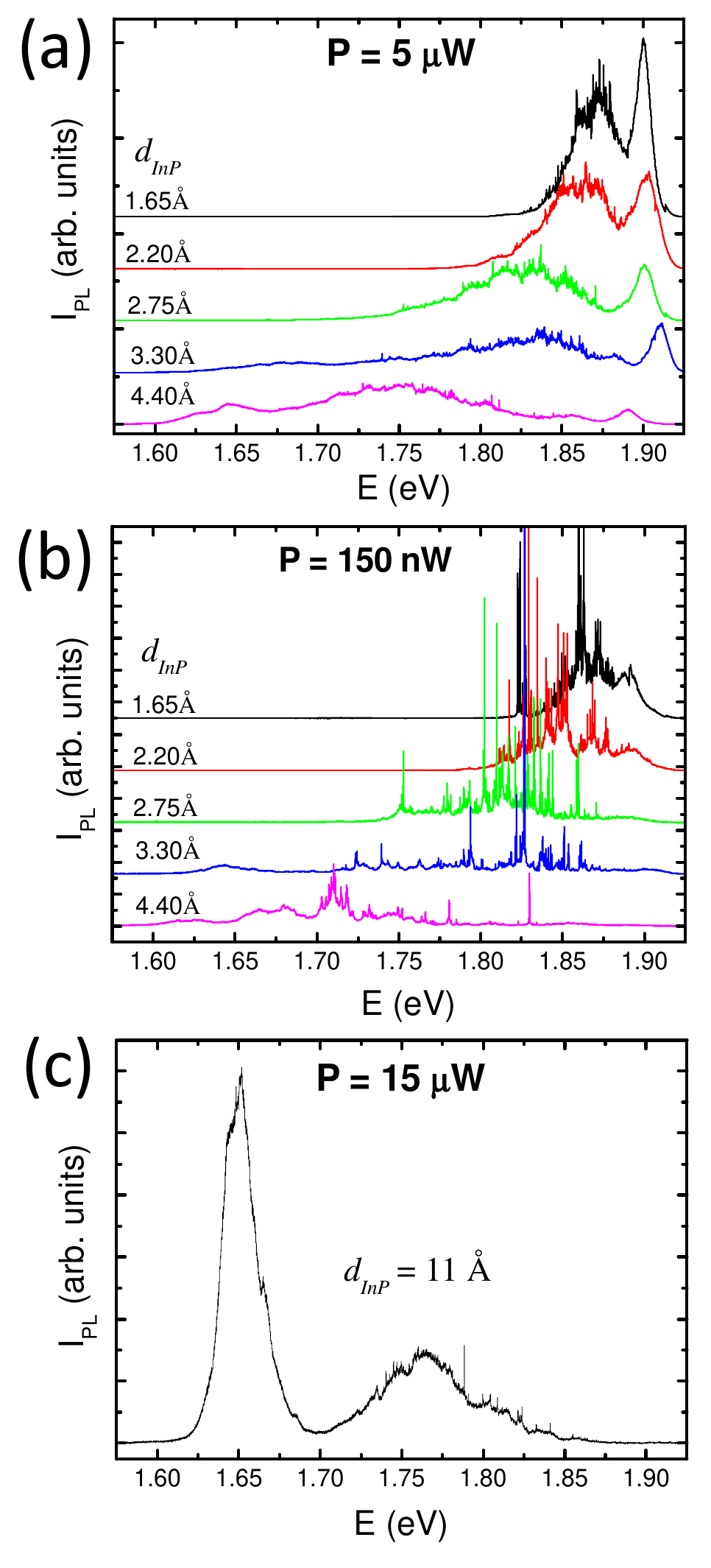}
\caption{$\mu$PL spectra of $\mathrm{InP/GaInP}$ QD ensembles measured at temperature 10K with a HeNe laser ($E_{HeNe}$=1.96~eV). Spectra for samples with $d_{InP}$ varying from $1.65$\AA\ to $4.4$~\AA~ are shown for laser excitation powers: (a) $P=5\mu\mathrm{W}$ and (b) $P=0.15{\mu}\mathrm{W}$. (c) Spectrum of a QD ensemble grown with $d_{InP}$=11\AA~ measured at $P=15{\mu}\mathrm{W}$.}
\label{fig1}
\end{figure}

The optical measurements were carried out using a micro-photoluminescence ($\mu$PL) set-up with  2 $\mu$m spatial resolution for a bare wafer or $\approx 1 \mu$m resolution defined by apertures in the opaque mask deposited on the sample surfaces. PL was analyzed using a 1~m double spectrometer with a charge coupled device (CCD). Photocurrent (PC) measurements were performed via resonant excitation with commercial current/temperature tunable laser-diodes and electrical detection with a picoammeter added to the $\mu$PL set-up circuit. All measurements were carried out at 10 K.

\section{III. Results}
\subsection{A. Ensemble characterization}

Figure~\ref{fig1} presents $\mu$PL spectra of unmasked samples obtained in the growth optimization procedure. Figure~\ref{fig1}(a) demonstrates that by varying the InP deposition thickness ($d_{InP}$) it is possible to control the size distribution of the InP QDs. As we observe, by changing $d_{InP}$ from 1.65 to 4.4~\AA, the center of the QD emission band can be shifted from 1.87 to 1.75~eV (the peak around $1.9~\mathrm{eV}$ corresponds to the $\mathrm{GaInP}$ barrier emission). This transition is markedly more gradual with deposition thickness than the one observed in the molecular beam epitaxy (MBE) growth of widely studied $\mathrm{InGaAs/GaAs}$ QDs\cite{Xie,qbit_book}. Importantly, this procedure also allows the formation of high densities of large QDs to be avoided. The large dots are formed for higher values of $d_{InP}$, which is illustrated in Fig.~\ref{fig1}(c), where, for $d_{InP}=11$\AA, we observe a pronounced multi-modal size distribution which is characterized by two broad PL peaks: a weaker peak at $1.765~\mathrm{eV}$ and a pronounced band at $1.65~\mathrm{eV}$, which correspond to small and large QD size distributions~\cite{Schulz,Persson,Hessman}, respectively.

The low excitation power spectra shown in Fig.~\ref{fig1}(b) demonstrate that the optimum conditions for the growth of low densities of small QDs are obtained for $d_{InP}$ values in the range of 2.75~\AA~ and 3.3~\AA. For these two values, we observe a relatively small number of individual QD PL emission lines, corresponding to QDs densities of $8\mathrm{x}10^{8}\mathrm{cm^{-2}}$ and $1\mathrm{x}10^{9}\mathrm{cm^{-2}}$, respectively, which are similar to the ones obtained by MBE growth~\cite{Ugur}. We note that the observed range of $d_{InP}$ (from 2.7~\AA~ to 3.3~\AA), which leads to growth of suitable samples at our $\mathrm{InP}$ deposition rate of $1.1$\AA$\mathrm{/s}$, is equivalent to $0.2$ atomic monolayers (ML). This range is large in comparison to the mechanical growth-control time, thus resulting in very reproducible growth confirmed in our further growth experiments. The variation of the dot density by a factor 1.25 in this range of $d_{InP}$  is up to a factor of 2 smaller than in MBE growth of $\mathrm{InP/GaInP}$ (or $\mathrm{InGaAs/GaAs}$) QDs for a similar range of deposition thicknesses and dot density around $10^9\mathrm{cm^{-2}}$~\cite{Xie,Ugur}. In this way, the MOVPE method discussed here offers a robust and well controlled method for fabrication of QD structures with ideal densities for individual QD studies, thus providing a suitable alternative to $\mathrm{InGaAs}$ structures grown by MBE.

\subsection{B. Single dot properties}

\begin{figure}[t]
\includegraphics[width=7.5cm]{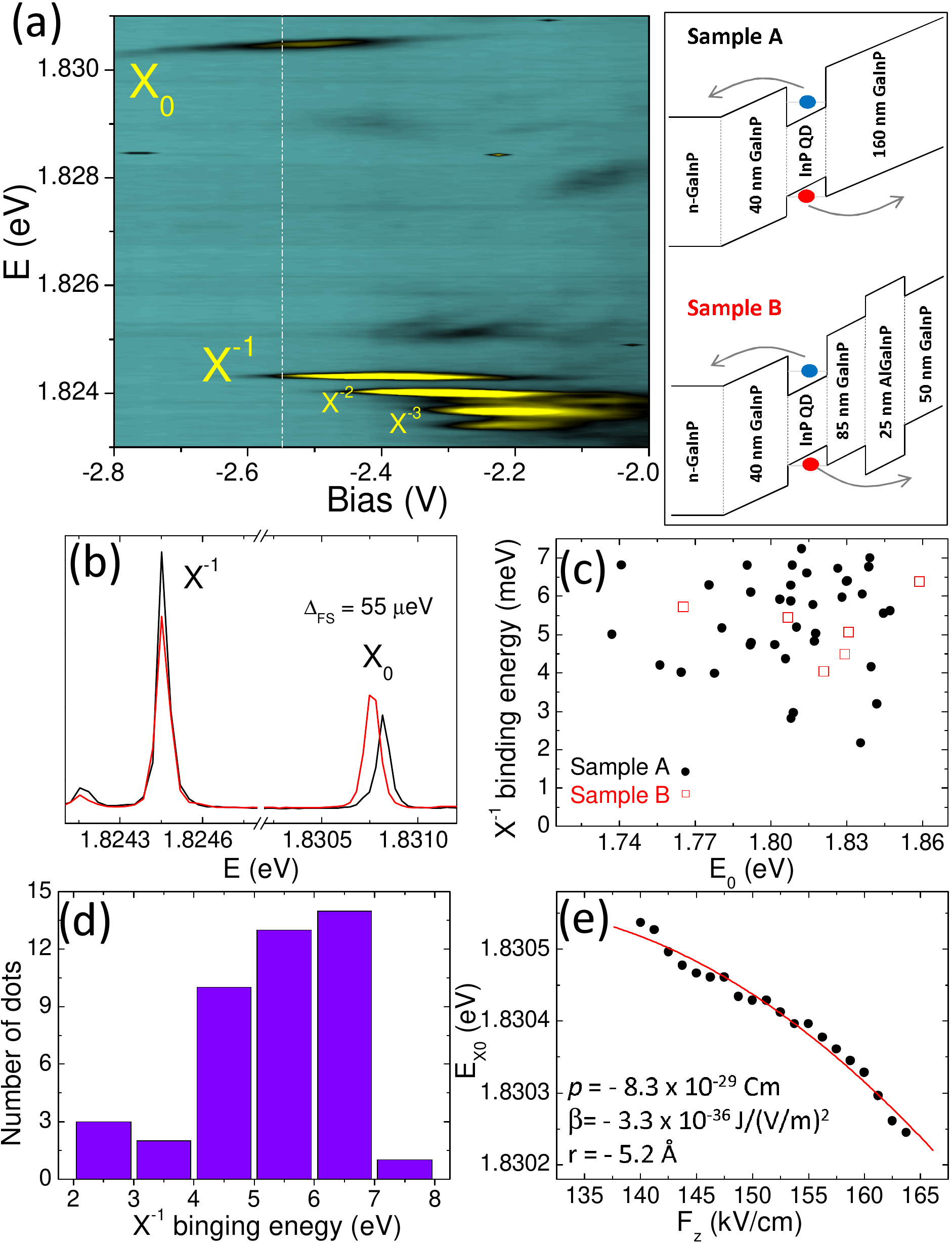}
   \caption{\label{Fig2} (a) Single QD $\mu$PL measured in sample B as a function of the bias applied between the \emph{n} and Schottky contacts. The inset illustrates Schottky-diode layer structure of samples A and B under reverse bias condition. (b) Linear polarization resolved PL measured at -2.55~V for the $X_0$ and $X^{-1}$ lines shown on (a). The black and red lines represent polarization
parallel to the [110] and $[1\overline{1}0]$ crystallographic directions, respectively. (c) Negatively charged exciton (X$^{-1}$) binding energy for samples A (black dots) and B (empty squares) against $E_0$, the energy of the neutral exciton at zero field. (d) Distribution of the X$^{-1}$ binding energy. (e) Neutral-exciton energy $E_{X0}$ as a function of the applied electric field $F_z$. Inset shows the values of $p$ and $\beta$ obtained from the fit to the data (solid curve).}
\end{figure}

After growth optimization, which led to identification of the optimum $d_{InP} \approx$~3\AA, samples A and B were grown as described in section II. The samples were processed in diodes with the top surfaces covered with a thin semi-transparent Ti layer and opaque Au-film contacts, where 1~$\mu$m apertures were open for optical access to the dots. Figure~\ref{Fig2}(a) shows an example of a bias dependence of the $\mu$PL spectrum of a single QD in sample B measured at excitation energy and power of 1.90~eV and 40~$\mu$W, respectively. For high negative bias (reverse bias) occupancy of the dot is low due to the high electron-hole tunneling rates. At V~$\approx$~-2.8~V a single emission line appears (marked as $X_0$). As the reverse bias is decreased, a second line appears on the low-energy side of $X_0$ separated by 6.32~meV. We attribute these two lines to the neutral ($X_0$) and singly-charged ($X^{-1}$) exciton states of the QD. This is confirmed by the cross-polarized linear PL detection measurements shown in Fig.~\ref{Fig2}(b). As expected, $X_0$ shows a fine-structure energy splitting ($\Delta_{FS}$=55~$\mu$eV), while $X^{-1}$ is insensitive to polarization of the detection because the electron-hole exchange interaction is  zero due to the presence of a second electron in the QD~\cite{Sheffield1,Sheffield2,Sheffield3,Nowak}. In Fig.~\ref{Fig2}(b), $X_0$ and $X^{-1}$ emission linewidths are 74 and 56~$\mu$eV, respectively, which are typical values observed for all QDs on samples A and B.

Figure~\ref{Fig2}(c) presents the $X^{-1}$ binding energies obtained for a large number of single QDs measured in samples A (dots) and B (empty squares) as a function of $E_0$, the $X_0$ energy at zero electric field (obtained from the fit of the QD Stark shift, as explained below). The binding energy is found not to depend on the confinement energy. The distribution presented in Fig.~\ref{Fig2}(d) shows that most of the QDs have binding energies between 4 and 7~meV, similar to what has been reported for InGaAs/GaAs based QDs~\cite{Warburton,Finley01,Warburton2}. This can be attributed to the similarity of effective masses and dielectric constant of both systems, which also should lead to similar values for biexciton binding energies~\cite{Wimmer}.

Note also in Fig.~\ref{Fig2}(a) that, as reverse bias continues to be decreased, electrons tunnel from the back \emph{n}-type contact into the QD, thus leading to the observation of more negatively charged exciton complexes, namely the $X^{-2}$ and $X^{-3}$~\cite{Warburton,Finley01}. In general, besides the $X_0$ emission, the  PL lines observed in the bias dependence measurements on samples A and B were predominantly due to negatively charged multi-exciton complexes. In sample B, the presence of the hole blocking barrier was expected to favour the formation of excitonic complexes with higher number of positive particles, i.e., positively charged excitons ($X^+$) and/or biexcitons ($XX$). However, emission of the $X^+$ was not identified in any of the samples, which indicates that, even with the hole blocking barrier, the tunneling times for holes in InP/$\mathrm{Ga_{0.5}In_{0.5}P}$ structures are short even at moderate electric fields. This agrees with calculations by Wimmer [19], where weak confinement for holes has been predicted in InP/$\mathrm{Ga_{0.5}In_{0.5}P}$ dots. Only a small number of QDs in sample B showed $XX$ emission lines that could be identified by measuring linearly-polarized PL. Nevertheless, precise determination of the biexciton binding energies was difficult because of the high spectral density of QD peaks at the low reverse bias where such lines start to be observed.

Figure~\ref{Fig2}(e) shows the emission energy ($E_{X0}$) of $X_0$ presented in Fig.~\ref{Fig2}(a) as a function of the applied electric field $F_z$, which is calculated by taking into account the Schottky barrier potential measured experimentally ($\approx$~0.5~V) and the separation between Ohmic and Schottky contacts (see inset Fig.~\ref{Fig2}). The solid line is a fit with the equation $E_{X0} = E_0 - pF_z + \beta F^2_z$, where $E_0$ is the energy at $F_z$=0, $p$ is the QD permanent dipole moment, and $\beta$ is the exciton polarizability \cite{Warburton2,Fry}. The values obtained for $p$ and $\beta$ are shown in the inset. From $p$ we extract an electron-hole separation $r = p/e$ = -5.2~\AA. The negative sign obtained for $r$ reflects the permanent dipole orientation at zero field for this particular QD: the electron is more localized in the direction of the apex and the hole in the direction of the base of the QD. The hole wavefunction located below that of the electron  has previously been inferred from PL bias dependence  measurements in InP QD ensembles~\cite{Hessman}. However, this is not always true at the single QD level because the electron-hole wavefunction alignment along \textbf{z}-direction is sensitive to the specific confinement characteristics of each individual dot. This property is discussed in detail in the following subsection.

\subsection{C. Exciton wavefunction}

\begin{figure}[b]
\includegraphics[width=7.3cm, clip]{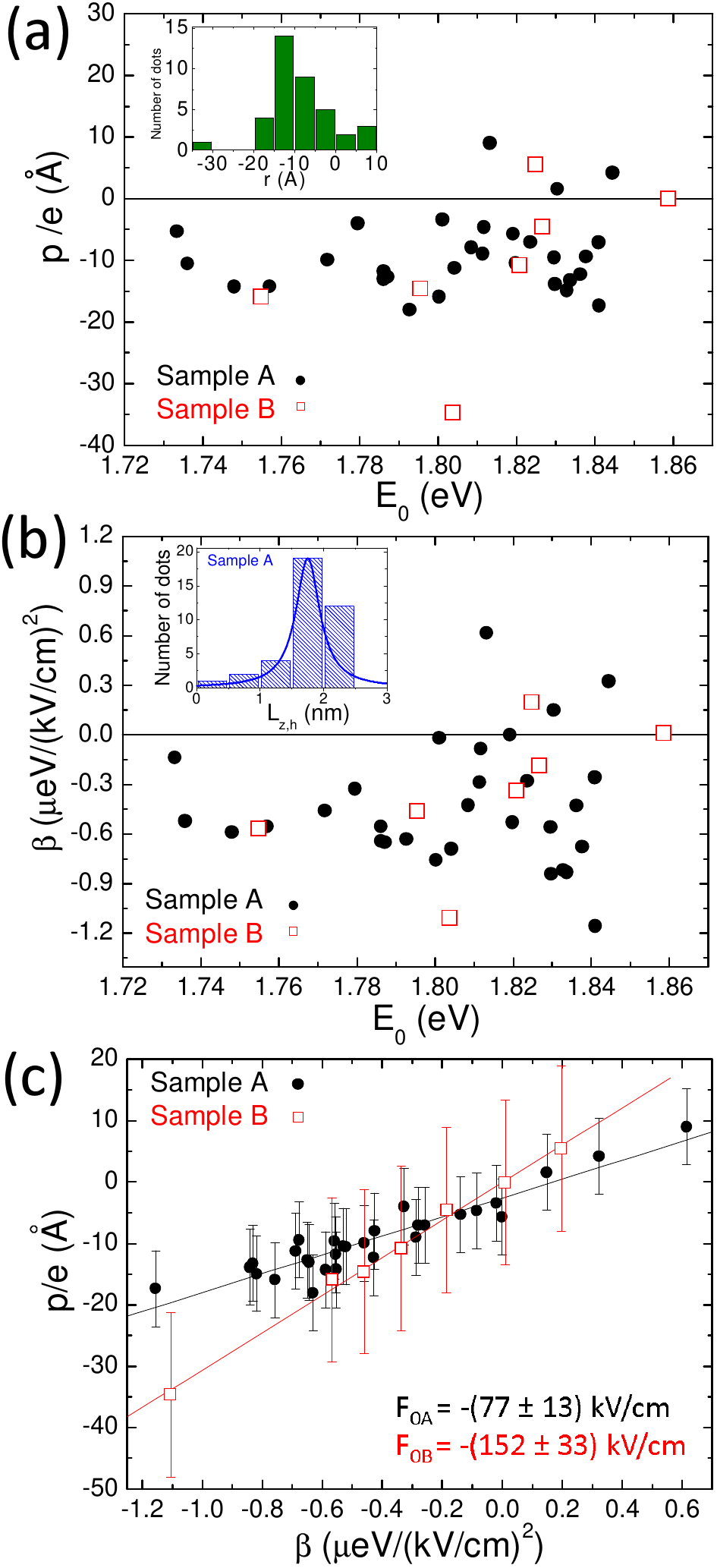}
   \caption{\label{Fig3b} (a) Permanent dipole moment $p$ and (b) polarizability $\beta$ plotted against the neutral exciton energy $E_0$ for samples A (dots) and B (empty squares). In (a), the inset shows the distribution of the electron-hole separation $r$. In (b), the inset shows the extent of the hole wavefunction along \textbf{z} direction ($L_{h,z}$) for sample A. (c) Permanent dipole moment $p$ against polarizability $\beta$ for samples A and B. Solid lines are linear fits to the data.}
\end{figure}

The dependence of the permanent dipole moment $p$ on $E_0$ for a large number of QDs in samples A (dots) and B (empty squares) is shown in Fig.~\ref{Fig3b}(a). We find that $p$, which is normally sensitive to the QD height and In concentration~\cite{Warburton2,Finley}, does not depend on the confinement energy in a range of approximately 150~meV. Besides this, the distribution of electron-hole separation $r$ presented on the inset shows that for the majority of the QDs the hole wavefunction is localized below that of the electron. Such a result is consistent with what is expected for strained QDs with a weak gradient of In distribution, for which a minimum of energy for holes is created at the QD base  due to the higher strain at the interface with the barrier~\cite{Pryor,Grundmann}.

However, as also observed in Fig.~\ref{Fig3b}(a), some QDs ($\approx$~13\%) show positive values for the permanent dipole moment $p$. As in the case of InGaAs/GaAs based systems, the occurrence of positive values for $p$ is possibly related to the presence of a positive In gradient from base to apex of the QDs~\cite{Fry,Barker,Finley}. This gradient, caused by the substitution of In by Ga atoms close to the interface with the barrier, ensures a smaller strain at the base with respect to the apex, thus inverting the natural dipole moment orientation of the system. NMR measurements by Chekhovich. \emph{et. al} (\emph{unpublished}) indicate occurrence between 10 and 15$\%$ of Ga in such nominally InP QDs, which can make some dots subject to composition inhomogeneities. The low occurrence of QDs with positive dipole (and the smaller magnitudes of such dipoles) is probably associated with the low mobility of Ga during growth since the QDs were grown at a lower temperature than the bottom GaInP barrier.

Figure~\ref{Fig3b}(b) presents the QD polarizability $\beta$ plotted as a function of $E_0$ for samples A (dots) and B (empty squares). Comparing distribution of $p$ and $\beta$ with $E_0$ in Figs.~\ref{Fig3b}(a) and (b), we can observe that there is a correspondence between the values of the QD permanent dipole moment and its polarizability. Such relationship provides information about the exciton wavefunction extension in the QD plane. This is clearly illustrated in Fig.~\ref{Fig3b}(c), where the values of the permanent dipole moments are displayed as function of the polarizability for both samples. In the harmonic confinement potential approximation, the presence of a permanent dipole moment can be associated with a built-in electric field $F_0$ along the growth direction. It can be easily shown that the ratio between dipole moment and polarizability characterizes this field $p/\beta = -2F_0$~\cite{Warburton2}. By fitting the experimental data with this equation, we obtain the fields $F_{0A} = -(77\pm13)$~kV/cm for sample A and $F_{0B} = -(152\pm33)$~kV/cm for sample B. The presence of an approximately constant built-in field for the two ensembles of QDs allows for a classical interpretation of the electron and hole wavefunctions as representing the two plates of a circular capacitor~\cite{Warburton2}. In that case, $F_0$ depends only on the area ($A$) of the capacitor $F_0=e/A\varepsilon_0\varepsilon_r$, not on the distance between the plates. Here, $\varepsilon_r=$~12.6 is the InP dielectric constant~\cite{Landolt}. This relationship allows us to estimate, independent of the size of the permanent dipole moment, the lateral extension of the excitonic wavefunction by assuming it to be determined by the area $A$. From $F_{0A}$ and $F_{0B}$ we obtain $a_A = 7.7$~nm and $a_B = 5.5$~nm for the average excitonic radius encountered in samples A and B, respectively. By comparing with the calculations of Wimmer \emph{et.al}~\cite{Wimmer}, the excitonic radius obtained experimentally from us should correspond to QDs with diameters around 20 and 30~nm.

Furthermore, for InP/$\mathrm{Ga_{0.5}In_{0.5}P}$ QDs, the statistics on QD polarizability provides information specifically about the hole wavefunction. Assuming a parabolic confinement in the vertical direction $\mathbf{z}$ for electrons and holes, the Stark shift of the states is given by $\Delta E = -(e^2/2\hbar^2)(m_h L^4_{h,z} - m_e L^4_{e,z})F^2_z$, so that the polarizability depends on the electron (hole) effective mass $m_e$ ($m_h$) and spatial extent of the wavefunction along confinement direction $L_{e,z}$ ($L_{h,z}$)~\cite{Barker,Warburton2}. However, besides the fact that $m_h > m_e$, for InP/$\mathrm{Ga_{0.5}In_{0.5}P}$ structures the confinement energy for holes along the vertical direction is expected to be much smaller then the one for electrons~\cite{Wimmer}. This implies that the hole wavefunction is expected to be more delocalized than the electron one ($L_{h,z} > L_{e,z}$), in contrast to the case of InGaAs/GaAs system~\cite{Warburton3}. The polarizability measured in our QDs is, therefore, characterized by the contribution of holes and can be written as $\beta \approx (e^2/2\hbar^2)m_h L^4_{h,z}$. Using the values obtained from the statistics performed for $\beta$ and assuming hole effective masse values for InP/$\mathrm{Ga_{0.5}In_{0.5}P}$ given by Wimmer~\cite{Wimmer}, we plot on the inset of Fig.~\ref{Fig3b}(b) the distribution for the extent of the hole wavefunction along the confinement direction $L_{h,z}$ for the dots probed in sample A. As we observe, the hole wavefunction extension is on average around 2~nm, which, comparing to Fig.~\ref{Fig3b}(a), corresponds to QDs with permanent dipoles $p \approx -10~{\AA}$.

This analysis allows to make more specific conclusions about the nature of the electron-hole alignment in the studied dots. Since electron is under a higher confinement regime the sign of the dipole for InP/$\mathrm{Ga_{0.5}In_{0.5}P}$ QDs is mainly determined by the position of the center of the hole wavefunction, which is more sensitive to the type of confinement added by the strain. As we observe experimentally, the majority of the QDs are characterized by an electron-above-hole alignment which occurs as a consequence of a high strain at the QD base but also indicates a more homogeneous distribution of In along its height. On the other hand, the presence of a higher concentration of Ga at the base of a small number of QDs relieves the strain at the interface and enhances the In gradient along the QD height, thus contributing to the appearance of smaller positive dipole moments. The height and strain distribution in the QDs, however, do not affect the in-plane extent of the exciton wavefunction, as demonstrated by the linear relationship between $p$ and $\beta$ shown in Fig.~\ref{Fig3b}(c).

\subsection{D. Photocurrent of single dots}

\begin{figure}[t]
\includegraphics[width=7.5cm]{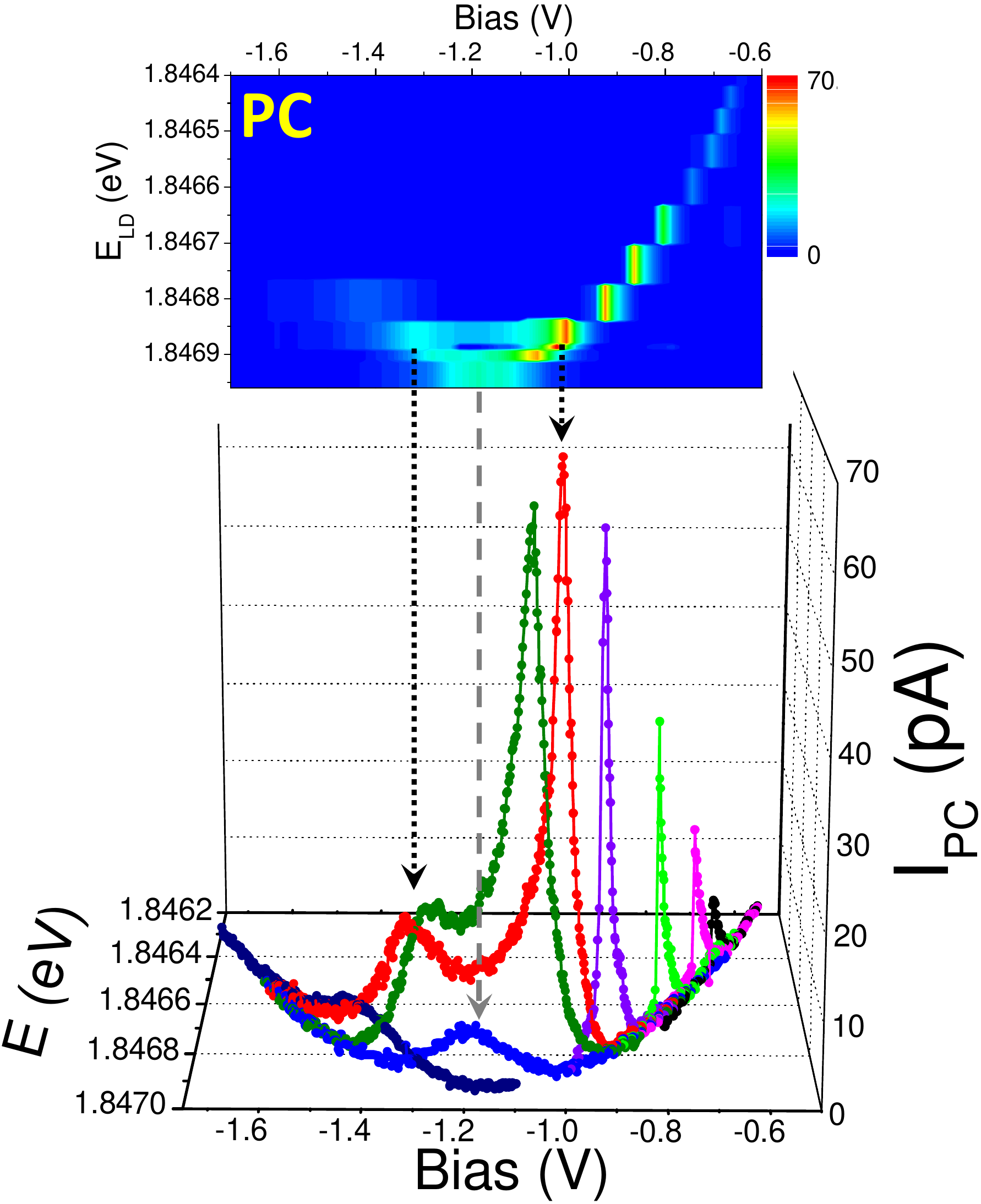}
   \caption{\label{Fig4} Photocurrent (PC) of a neutral exciton state $X_0$ measured in a single InP QD. The central panel shows a 3D-plot with PC spectra. Each PC spectrum is measured at a fixed excitation-laser energy $E_{LD}$ by tuning the $X_0$ exciton energy using bias (see text for details). The top panel shows a 2D-plot with amplitudes of the PC signal measured  for different laser excitation energies $E_{LD}$, where the parabolic Stark-shift clearly observed. Arrows connect the parts of the parabola in the 2D-plot and the corresponding PC peaks observed in the 3D-plot.}
\end{figure}

Figure~\ref{Fig4} presents resonant-excitation experiments performed on a single InP QD using photocurrent (PC) techniques. The main panel of the figure shows a 3D-plot of PC spectra measured for the neutral exciton ($X_0$) in one of the QDs in sample A. Each PC curve is measured by fixing the energy ($E_{LD}$) of a single-mode laser-diode and tuning the QD neutral exciton energy through the resonance with $E_{LD}$ by changing the applied bias. The upper panel in Fig.~\ref{Fig4} shows the measured PC amplitude for different values of $E_{LD}$ as the QD energy is changed by the applied bias. A characteristic parabolic Stark shift is observed as a function of bias. By setting $E_{LD}~=$~1.84695~eV, a PC curve with one peak is measured at approximately -1.2~V, corresponding to the maximum energy in the parabola. The vertical gray arrow points at  the PC spectrum in the 3D-plot corresponding to this $E_{LD}$. As $E_{LD}$ is decreased, first two PC peaks are observed at different bias. The corresponding points on the Stark-shift parabola and the PC maxima in the 3D-plot are linked by black arrows on the graph. A significant broadening of the PC peak at high bias is observed, arising due to the fast tunneling of the carriers from the dot in high electric field \cite{Oulton,Makhonin}. As $E_{LD}$ is decreased further, only one relatively sharp PC peak at low bias is clearly observed.

Note that the bias regime where PC is measured overlaps with that of PL. This PC property is observed in all QDs measured in samples A and B. This is contrary to InGaAs QDs, where PC is normally observed at higher reverse bias regime as compared to PL emission~\cite{Oulton,Makhonin}, as the carrier tunneling rates have to be higher than the electron-hole recombination rate. The low bias regime where PC is observed for InP QDs is, therefore, possibly related to the high hole tunneling rate arising as a consequence of the weak hole confinement.

To conclude on this section, Fig.~\ref{Fig4} clearly demonstrates that single InP QD states can be directly addressed by resonant excitation and detected electrically (as previously reported for InGaAs QDs only), thus opening the way for more sophisticated experiments involving resonant manipulation of electron, hole and nuclear spins in single dots employing bias control \cite{Ramsay1,Ramsay2,Makhonin,Kloeffel}.

\section{IV. Conclusions}

In summary, by realizing MOVPE growth of low density $\mathrm{InP/GaInP}$ QDs, we have overcome the major hurdle of the presence of high densities of large QDs in this system. We achieve a reproducible and smooth transition in QD size distribution and density by varying nominal InP deposition thickness. This has allowed precise control of the charge state in individual InP QDs by application of vertical electric fields using Schottky devices: by tuning the applied bias we demonstrate regimes where neutral ($X_0$) and negatively charged ($X^{-1}$) excitons can be clearly observed. $X^{-1}$ binding energies are shown to range from 4 to 7~meV, similar to InGaAs/GaAs QDs. Systematic studies of the exciton permanent dipole moment and polarizability in a large number of individual InP QDs allows for characterization of the exciton wavefunction in such system. We argue that due to a relatively higher confinement for electrons, the sign of the exciton permanent dipole moment is mainly determined by the position of the hole wavefunction along the growth direction. This provides insight into the QD composition and strain distribution. Moreover, from the relationship between dipole moment and polarizability, we show that the lateral extent of the exciton wavefunction varies very little from dot to dot in the same sample. We obtain an average in-plane exciton radius of 7.7 and 5.5~nm for QDs probed in two different samples. We also demonstrate photocurrent techniques, allowing for resonant manipulation and electrical detection of excitons in single $\mathrm{InP/GaInP}$ QDs.

\section{Acknowledgments}

We thank O. Gazzano for participation in initial experiments and M. Hugues for fruitful discussions. This work has been supported by the EPSRC Programme Grant EP/G601642/1, the Royal Society and ITN Spin-Optronics. J. Puebla has been supported by CONACYT-Mexico Doctoral Scholarship.

\end{document}